# The Role of Correlation in the Operation of Quantum-dot Cellular Automata

**Géza Tóth and Craig S. Lent**
*Department of Electrical Engineering*
*University of Notre Dame*
*Notre Dame, IN 46556*
*e-mail: Geza.Toth.17@nd.edu, Craig.S.Lent.1@nd.edu*

*[*]Also Neuromorphic Information Technology Graduate Center*
*Kende-u. 13, Budapest H-1111, Hungary*
*Present address: Géza Tóth, Theoretical Physics,*
*University of Oxford, 1 Keble Road, Oxford OX1 3NP, UK*

*Author to whom correspondence should be addressed:*
*Géza Tóth, e-mail: Geza.Toth.17@nd.edu,*

## ABSTRACT

Quantum-dot Cellular Automata (QCA) may offer a viable alternative of traditional transistor-based technology at the nanoscale. When modeling a QCA circuit, the number of degrees of freedom necessary to describe the quantum mechanical state increases exponentially making modeling even modest size cell arrays difficult. The intercellular Hartree approximation largely reduces the number of state variables and still gives good results especially when the system remains near ground state. This suggests that large part of the correlation degrees of freedom are not essential from the point of view of the dynamics. In certain cases, however, such as for example the majority gate with unequal input legs, the Hartree approximation gives qualitatively wrong results. An intermediate model is constructed between the Hartree approximation and the exact model, based on the coherence vector formalism. By including correlation effects to a desired degree, it improves the results of the Hartree method and gives the approximate dynamics of the correlation terms. It also models the majority gate correctly. Beside QCA cell arrays, our findings are valid for Ising spin chains in transverse magnetic field, and can be straightforwardly generalized for coupled two-level systems with a more complicated Hamiltonian.

# I. Introduction

In recent years the development of integrated circuits has been essentially based on scaling down, that is, increasing the element density on the wafer. Scaling down of CMOS circuits, however, has its limits. Above a certain element density various physical phenomena, including quantum effects, conspire to make transistor operation difficult if not impossible. If a new technology is to be created for devices of nanometer scale, new design principles are necessary. One promising approach is to move to a transistor-less cellular architecture based on interacting quantum dots, Quantum-dot Cellular Automata (QCA, [1-7]).

The QCA paradigm arose in the context of electrostatically coupled quantum dots. A QCA cell consists of four (or five) such dots arranged in a square pattern. Information is encoded in the arrangement of charge (i. e., two extra electrons) within the cell. When a cell is switched, these electrons tunnel through interdot barriers to neighboring dots inside the cell.

The physical descriptions of the two limiting cases are the semi-classical QCA dynamics [8, 10] and the quantum QCA dynamics[4]. The first is used to model the metallic-island implementation of the QCA circuits where the dots contain many electrons (however, only the two extra electrons tunnel to neighboring dots when being switched) and the circuit is described in terms of classical quantities as charging energies, capacitance, etc. The quantum QCA dynamics is used to model cells which can be considered a quasi two-state system with a coherent time evolution. (Decoherence can also be included in the model.)

After developing the basic logic gates the theory has been extended to large arrays of devices and computer architecture questions. A key advance was the realization that by periodically modulating the inter-dot barriers, clocked control of QCA circuitry could be accomplished. The modulation could be done at a rate which is slow compared to inter-dot tunneling times, thereby keeping the switching cells very near the instantaneous ground state. This quasi-adiabatic switching [4] paradigm has proven very fruitful. Quasi-adiabatic clocking permits both logic and addressable memory to be realized within the QCA framework. It allows a pipe-lining of computational operations.

The ability of modeling large cell arrays is crucial for the development of complex QCA circuits. Unfortunately, the number of quantum degrees of freedom increases exponentially with the system size. Using the Hartree approximation reduces the number of state variables drastically and it can still give quantitatively good results in many cases. Thus even ignoring many quantum degrees of freedom, the dynamics obtained from the model remains close to the "exact" dynamics obtained from the many-body Schrödinger equation. In other cases, the Hartree method can give quantitatively wrong results.

The intercellular Hartree approximation [7] can reduce the number of state variables since it neglects all the correlations. In general, the correlation of two quantities, $A$ and $B$ can be defined as [9]

$$C_{AB} = \langle AB \rangle - \langle A \rangle \langle B \rangle . \tag{1}$$

It was shown [10] that in the case of quasi-classical QCA dynamics cell correlation plays an important role and assuming $C_{AB}$=0 will lead to wrong results. In this paper the role of correlations in the quantum QCA dynamics is examined. A model is proposed which makes it possible to include as much quantum correlation degrees of freedom as necessary for obtaining the correct dynamics. The model will be tested on two examples: a cell line and a majority gate with unequal input legs. In the first case the dynamics is quantitatively improved with respect to the model using the Hartree intercellular approximation. In the second case the Hartree method gives even qualitatively wrong results. Our model gives the correct results by including correlation effects.

In Sec. II the Quantum-dot Cellular automata with quasi-adiabatic switching is reviewed. In Sec. III. the coherence vector formalism is introduced. Sec. IV. describes a model that makes it possible to neglect higher order correlations. In Sec. V. simulation examples are shown to compare the results of the exact and the approximate method.

## II. Quasi-adiabatic switching with Quantum-dot Cellular Automata

The QCA cell consists of four quantum dots as shown in Fig. 1(a). Tunneling is possible between the neighboring dots as denoted by lines in the picture. Due to Coulombic repulsion the two electrons occupy antipodal sites as shown in Fig. 1(b). These two states correspond to charge polarization +1 and -1, respectively, with intermediate polarization interpolating between the two.

In Fig. 1(c) a two cell arrangement is shown to illustrate the cell-to-cell interaction. Cell 1 is a driver cell whose polarization takes the range -1 to 1. It is also shown, how the polarization of cell 2 changes for different values of the driver cell polarization. It can be seen, that even if the polarization of the driver cell 1 is changing gradually from -1 to +1, the polarization of cell 2 changes abruptly from -1 to +1. This *nonlinearity* is also present in digital circuits where it helps to correct deviations in signal level: even if the input of a logical gate is slightly out of the range of valid “0” and “1” voltage levels, the output will be correct. In the case of the QCA cells it causes that cell 2 will be saturated (with polarization close to -1 or +1) even if cell 1 was far from saturation.

A one-dimensional array of cells[4] can be used to transfer the polarization of the driver at one end of the cell line to the other end of the line. Thus the cell line plays the role of the wire in QCA circuits. Moreover, any logical gates (majority gate, AND, OR) can also be implemented, and using these as basic building elements, any logical circuit can be realized[5].

In this paradigm of ground state computing, the solution of the problem has been mapped onto the ground state of the array. However, if the inputs are switched *abruptly*, it is not guaranteed that the QCA array really settles in the ground state, i.e., in the global energy minimum state. It is also possible, that eventually it settles in a *metastable* state because the trajectory followed by the array during the resulting transient is not well controlled.

This problem can be solved by quasi-adiabatic switching [4] of the QCA array, as shown schematically in Fig. 2. Quasi-adiabatic switching has the following steps: (1) before applying the new input, the height of the interdot barriers is lowered thus the cells have no more two distinct polarization states, P=+1 and P=-1. (2) Then the new input can be given to the array. (3) While raising the barrier height, the QCA array will settle in its new ground state.

The quasi-adiabatic switching is based on the adiabatic theorem, which states that if the change of the Hamiltonian is gradual enough and the system is initially in ground state then it will stay in ground state throughout the whole switching process. Because the system is minimally excited from the ground state, dissipation to the environment is very small. On the other hand, to maintain quasi-adiabaticity the time over which the barrier height is modulated must be long compared to the tunneling time through the barrier. Typically a factor of 10 reduces the non-adiabaticity to very small levels.

## III. Coherence vector formalism applied for the Quantum-dot Cellular Automata

The Hamiltonian for a QCA circuit modelled as coupled two-state systems [3] is:

$$\hat{H} = -\gamma\sum_{i=1}^{N}\hat{\sigma}_x(i) - \sum_{i=1}^{N-1}\sum_{j=i+1}^{N}\frac{E_{ij}}{2}\hat{\sigma}_z(i)\hat{\sigma}_z(j) + \frac{E_0}{2}\sum_{i=1}^{N}\hat{\sigma}_z(i)P_{driver}(i) \quad , \quad (2)$$

where $E_{ij}$ is the electrostatic coupling between the $i^{th}$ and the $j^{th}$ cell, and $\gamma$ is the tunneling energy. The first term describes the intracell tunneling between the two basis states. The second term describes the electrostatic coupling between neighbors. The third term describes coupling to driver cells. For those cells which do not have a driver cell as a neighbor $P_{driver}(i)=0$.

If the interdot tunneling barriers in the cells are high and the tunneling rate is very low (zero), then the $\gamma$ tunneling energy is zero. If the interdot tunneling barriers in the cells are low and the tunneling rate is high, then $\gamma$ is large. The tunneling barriers of the cells are connected to electrodes and their heights is controlled externally by voltage sources.

For a cell line the nearest neighbor couplings can be given by $E_{i,i+1}=E_0$ while all the other $E_{ij}$'s are zero. In this case $E_{i,i+1}$ is the cost in electrostatic energy for two cells being oppositely polarized.

The polarization of the $k^{th}$ cell can be interpreted as the expectation value of the $\hat{\sigma}_z(k)$ Pauli spin matrix:

$$P(k) = -\langle\hat{\sigma}_z(k)\rangle. \quad (3)$$

With the negative sign we follow the convention of Ref. [15] choosing the sign of the Pauli spin matrices:

$$\hat{\sigma}_x = \begin{bmatrix} 0 & 1 \\ 1 & 0 \end{bmatrix}, \hat{\sigma}_y = \begin{bmatrix} 0 & i \\ -i & 0 \end{bmatrix}, \text{ and } \hat{\sigma}_z = \begin{bmatrix} -1 & 0 \\ 0 & 1 \end{bmatrix}. \tag{4}$$

The dynamics of the cell line can be computed by the Liouville equation giving the time dependence of the density matrix. The density matrix of a system of $N$ cells has $2^N$x$2^N$ complex (=2x$2^{2N}$ real) elements. $2^{2N}$ +1 constraints are coming from the requirements of Hermiticity and unit trace leaving $s=2^{2N}-1$ real (i.e., not complex) degrees of freedom. Now the density matrix can be expressed as a linear combination of the $s$ generating operators of the special unitary group SU($2^N$):

$$\hat{\rho} = \frac{1}{2^N}\hat{1} + \frac{1}{2^N}\sum_{i=1}^{s} \Lambda_i \hat{\Lambda}_i . \tag{5}$$

where

$$\Lambda_i = \langle \hat{\Lambda}_i \rangle. \tag{6}$$

The $\hat{\Lambda}_i$ basis operators have the form:

$$\hat{\Lambda}_i = \hat{\lambda}_i^{(1)} \otimes \hat{\lambda}_i^{(2)} \otimes \ldots \otimes \hat{\lambda}_i^{(N)} . \tag{7}$$

where a term of the Kronecker product can be one of four single-cell operators:

$$\hat{\lambda}_i^{(k)} = \begin{cases} \hat{1} \\ \hat{\sigma}_x(k) \\ \hat{\sigma}_y(k) \\ \hat{\sigma}_z(k) \end{cases} . \tag{8}$$

Since choosing only $\hat{1}$'s is excluded, there are $s=4^N-1$ $\hat{\Lambda}_i$'s. (For example, $\hat{\sigma}_x(1)\hat{\sigma}_y(2)\hat{\sigma}_x(3)$, $\hat{\sigma}_y(1)\hat{\sigma}_z(3)$ and $\hat{\sigma}_z(1)$ are among the basis operators.)

In this paper the vector constructed from the $\Lambda_i$ coefficients of the (5) linear combination, the $\vec{\Lambda}$ coherence-vector[15], will be used for the state description instead of the density matrix. The elements of the $\vec{\Lambda}$ coherence vector are the expectation values of the $\hat{\Lambda}_i$ basis operators. The coherence vector can be partitioned into $\vec{\lambda}(i)$ single-cell coherence vectors, $\vec{K}(i, j)$ two-point, $\vec{K}(i, j, k)$ three-point etc., correlation vectors:

$$\vec{\Lambda} = \left[\vec{\lambda}(1)\ \vec{\lambda}(2)\ \dots\ \vec{K}(1,2)\ \vec{K}(1,3)\ \dots\ \vec{K}(1,2,3)\ \dots\right]^T . \tag{9}$$

The $\vec{\lambda}(i)$ single cell coherence vectors contain the expectation values of the $\hat{\sigma}_x(i)$, $\hat{\sigma}_y(i)$ and $\hat{\sigma}_z(i)$ single-cell basis operators. The $\vec{K}(i, j)$ two-point correlation vector has nine elements:

$$\vec{K}(i, j) = \left[K_{xx}\ K_{xy}\ K_{xz}\ K_{yx}\ K_{yy}\ K_{yz}\ K_{zx}\ K_{zy}\ K_{zz}\right]^T , \tag{10}$$

They are expectation values of two-cell basis operators:

$$K_{ab}(i, j) = \langle \hat{\sigma}_a(i)\hat{\sigma}_b(j)\rangle; \qquad a, b = x, y, z . \tag{11}$$

Similarly, the elements of the three-point correlations are expectation values of three-cell basis operators:

$$K_{abc}(i, j, k) = \langle \hat{\sigma}_a(i)\hat{\sigma}_b(j)\hat{\sigma}_c(k)\rangle; \qquad a, b = x, y, z . \tag{12}$$

The dynamics of the coherence vector elements can be obtained by first computing the dynamics of the basis operators in the Heisenberg picture and then taking the expectation values of both sides of the equations. The differential equation system is linear and has the form:

$$\hbar\frac{d}{dt}\vec{\Lambda} = \hat{\Omega}(t)\vec{\Lambda}, \tag{13}$$

where $\hat{\Omega}(t)$ is the time dependent coefficient matrix. Next the structure of the (13) differential equation system will be presented by giving explicit equations for the single cell coherence vector elements and two-point correlations.

The dynamics of a single cell coherence vector can be obtained as

$$\hbar\frac{d}{dt}\vec{\lambda}(i) = \hat{\Omega}_i\vec{\lambda}(i) + \sum_{j \in Nb(i)} E_{ij}\left[\langle\hat{\sigma}_y(i)\hat{\sigma}_z(j)\rangle \; -\langle\hat{\sigma}_x(i)\hat{\sigma}_z(j)\rangle \; 0\right]^T, \quad (14)$$

where

$$\hat{\Omega}_i = \begin{bmatrix} 0 & -E_0P_{driver}(i) & 0 \\ E_0P_{driver}(i) & 0 & 2\gamma \\ 0 & -2\gamma & 0 \end{bmatrix}, \quad (15)$$

and *Nb(i)* refers to the neighbors of the $i^{th}$ cell. The first term on the right hand side of (14) describes the precession of $\vec{\lambda}(i)$ around an axis determined by $P_{driver}(i)$ and $\gamma$. The second term with the sum is the coupling to the neighbors through two-point correlations.

The $K_{yz}(i, j) = \langle\hat{\sigma}_y(i)\hat{\sigma}_z(j)\rangle$ and $K_{xz}(i, j) = \langle\hat{\sigma}_x(i)\hat{\sigma}_z(j)\rangle$ terms are two-point correlation vector elements. The dynamics of the correlation vectors can be obtained as[11]:

$$\hbar\frac{d}{dt}\vec{K}(i, j) = (\hat{1} \otimes \hat{\Omega}_j + \hat{\Omega}_i \otimes \hat{1})\vec{K}(i, j) + \vec{C}_{ij}\{\vec{\lambda}(k), \vec{K}(l, m, n)\}. \quad (16)$$

The first term on the right hand side of (16) corresponds to the evolution of $\vec{K}(i, j)$ under the influence of $P_{driver}(i)$, $P_{driver}(j)$ and $\gamma$. The second term with $\hat{C}_{ij}$ is an expression consisting of coherence vector elements and three-point correlation vector elements.

Dynamical equations similar to (16) can be written for the three-point, four-point, etc. correlation vector elements. (They are not given here.) The complete set of these differential equations describes the dynamics of the multi-cell system equivalently to the dynamics given by the Liouville equation for the density matrix. If there is no decoherence and the system is in a pure state, these two are equivalent to the dynamics given by the Schrödinger equation with the many-cell Hamiltonian. We will refer to the model containing the whole set of differential equations for the coherence vector and correlation vector elements as the *exact* model in this paper.

The structure of the dynamical equations for the correlation vector elements is such that in the equation of $n^{th}$ order correlation vector elements we can find only $(n+1)^{th}$ and lower order correlations [12]. This provides a possibility to truncate the hierarchy of dynamical equations. Having formulas which approximate the $(n+1)^{th}$ order correlations with lower order correlations and substituting them in the equations of the $n^{th}$ order terms, all differential equations for the terms with order higher than $n$ can be neglected since these terms cannot be found in the equations of lower order correlation terms. The details of the truncation will be given in the next section.

Besides the correlation vector there are other quantities characterizing the intercell correlation. The *correlation vector proper* [13] for two cells has nine elements. They are defined as

$$M_{ab}(i,j) = \left\langle \left[\hat{\sigma}_a(i) - \langle\hat{\sigma}_a(i)\rangle\right]\left[\hat{\sigma}_b(j) - \langle\hat{\sigma}_b(j)\rangle\right]\right\rangle; \quad a,b = x,y,z \tag{17}$$

With coherence vector elements (17) can be rewritten as

$$M_{ab}(i,j) = K_{ab}(i,j) - \lambda_a(i)\lambda_b(j); \qquad a,b = x,y,z. \tag{18}$$

The elements of the correlation vector proper are all zero if there is no correlation between the cells or they are *uncorrelated*.

The higher order correlation vectors proper are defined similarly to (17). For example, an element of the three-point correlation vector proper can be given as

$$M_{abc}(i,j,k) = \left\langle \left[\hat{\sigma}_a(i) - \langle \hat{\sigma}_a(i) \rangle\right] \left[\hat{\sigma}_b(j) - \langle \hat{\sigma}_b(j) \rangle\right] \left[\hat{\sigma}_c(k) - \langle \hat{\sigma}_c(k) \rangle\right] \right\rangle;$$

$$a, b, c = x, y, z. \quad (19)$$

After some algebraic transformations one gets

$$M_{abc}(i,j,k) = K_{abc}(i,j,k) - K_{ab}(i,j)\lambda_c(k) - K_{ac}(i,k)\lambda_b(j) - K_{bc}(j,k)\lambda_a(i) + 2\lambda_a(i)\lambda_b(j)\lambda_c(k); \qquad a, b, c = x, y, z. \quad (20)$$

**The intercellular Hartree approximation**

It is possible to eliminate the correlation terms from (14) by assuming that the cells are uncorrelated, that is, the two-point correlation vectors proper are zero:

$$M_{ab}(i, i+1) = K_{ab}(i, i+1) - \lambda_a(i)\lambda_b(i+1) = 0. \quad (21)$$

Based on (21) the two-point correlation vector elements can be approximated with the multiplication of two coherence vector elements:

$$K_{ab}(i, i+1) \approx \lambda_a(i)\lambda_b(i+1). \quad (22)$$

Substituting (22) into (14) one obtains the dynamical equations for the coherence vectors as

$$\hbar \frac{d}{dt}\vec{\lambda}(i) = \tilde{\Omega}_i \vec{\lambda}(i), \quad (23)$$

where

$$\tilde{\Omega}_i = \begin{bmatrix} 0 & -\Sigma_i & 0 \\ \Sigma_i & 0 & 2\gamma \\ 0 & -2\gamma & 0 \end{bmatrix}, \text{ and} \quad (24)$$

$$\Sigma_i = \sum_{j \in Nb(i)} E_{ij} P(j) + E_0 P_{driver}(i) . \tag{25}$$

Here $\Sigma_i$ is the weighted sum of the polarizations of the neighbors ($P(k) = -\lambda_z(k)$).

The (23) dynamical equation can be written [15] in the form of

$$\hbar \frac{d}{dt} \vec{\lambda}(i) = \vec{\Gamma}(i) \times \vec{\lambda}(i) \quad , \tag{26}$$

where the cross denotes vector product and

$$\vec{\Gamma}(i) = \begin{bmatrix} -2\gamma & 0 & \Sigma_i \end{bmatrix}^T . \tag{27}$$

(26) describes the precession of $\vec{\lambda}(i)$ around $\vec{\Gamma}(i)$. The instantaneous ground state of (26) is

$$\vec{\lambda}_{ss}(i) = -\frac{\vec{\Gamma}_i}{\left|\vec{\Gamma}_i\right|} . \tag{28}$$

This approximation describes the state of the cell array by the single-cell coherence vectors only, using 3 real state variables for each cell. If there is no decoherence and the system is in a pure state then (26) is equivalent to the coupled Schrödinger equations[14]

$$i\hbar \frac{d\Psi_k}{dt} = \hat{H}_k \Psi_k ; \text{ k=1,2,...,N}, \tag{29}$$

where the single-cell Hamiltonians are

$$\hat{H}_k = -\gamma \hat{\sigma}_x(k) + \frac{\Sigma_k}{2} \hat{\sigma}_z(k) ; \quad \Sigma_i = \sum_{j \in Nb(i)} -E_{ij} \langle \hat{\sigma}_z(j) \rangle + E_0 P_{driver}(i) , \tag{30}$$

and the single-cell state functions are the superposition of the basis states

$$\Psi_k = \alpha_k |1\rangle + \beta_k |-1\rangle = \begin{bmatrix} \alpha_k \\ \beta_k \end{bmatrix} . \tag{31}$$

The state of the whole system can be constructed from the single-cell state vectors as $\Psi = \Psi_1 \otimes \Psi_2 \otimes \ldots \otimes \Psi_N$.

In [3] the (29) equations are used to model QCA lines where it is called intercellular Hartree approximation[14]. In this paper we will also call the model based on the (23-25) equations Hartree approximation or Hartree method.

# IV. Model neglecting higher order correlation

In a classical multi-particle system the number of degrees of freedom necessary for the state description increases linearly with the number of particles. A point-like particle can be described by its position and velocity. For $N$ particles, $N$ positions and $N$ velocities are required which gives $N$ times more degrees of freedom than for a single particle.

In a quantum mechanical system of $N$ QCA cells, the number of degrees of freedom are much larger than $N$ times the degrees of freedom of a single cell. The extra degrees of freedom come from the intercell *correlations.*The information necessary for a total description increases exponentially with the number of cells and makes it difficult to simulate even a modest size block of QCA cells. To describe $N$ coupled cells exactly, $2^{2N}$-1 variables are necessary for the coherence vector description.

The coherence vector description makes it possible to divide the state variables into groups corresponding to the state of the cells, and to the two-point, three-point, etc. correlations. A correlation term can be two-point, three-point, etc. or nearest neighbor, next-to-nearest neighbor, etc. This feature of the coherence vector description helps us to determine which correlation terms are important from the point of view of the dynamics and which can be neglected. Usually it is reasonable to assume that the further than nearest neighbor and higher order correlations play a less important role, thus they can be approximated by lower order correlations. Depending on which correlation terms are kept and which are neglected, models with different levels of approximations can be constructed which are intermediate between the Hartree approximation and the exact method.

Since it will be important later, we outline for a cell line the hierarchy of dynamical equations for the coherence and correlation vectors for the first three levels (coherence vector elements, two- and three-point correlations) in Table I. The figure shows which variables are on the right hand side for the dynamical equation for the single cell coherence vectors (graph I), the nearest neighbor two-point correlation vectors (graph II.a), the further-than-nearest neighbor two-point correlation vectors (graph II.b) and for the nearest neighbor three-point correlation vectors (graph III).

The Hartree approximation truncates the hierarchy at the dashed line in Table I keeping only graph I by removing the coupling to the two-point correlations indicated by the upper arrow. It assumes that the $M_{ab}(i, i+1)$ two-point correlation vector proper elements are zero (see (21)) and approximates the elements of the two-point correlation vectors with coherence vector elements using (22).

The first approximation, that is better than the Hartree method, can be obtained [16] by keeping only the single cell coherence vectors (graph I) and the two-point nearest neighbor correlations (graph II.a). The point of truncation is indicated by a dashed-dotted line in Table I. The truncation removes the coupling to the three-point correlations indicated by the lower arrow.

In order to do the truncation, a formula must be constructed to approximate the elements of the $\vec{K}(i, i+1, i+2)$ nearest neighbor three-point correlation vector with nearest neighbor two-point correlation vector and coherence vector elements. The approximation is based on the assumption that the (20) three-point correlation vector proper elements are zero:

$$K_{abc}(i, i+1, i+2) = \langle \hat{\sigma}_a(i)\hat{\sigma}_b(i+1)\hat{\sigma}_c(i+2)\rangle \approx K_{ab}(i, i+1)\lambda_c(i+2) + K_{bc}(i+1, i+2)\lambda_a(i) + K_{ac}(i, i+2)\lambda_b(i+1) - 2\lambda_a(i)\lambda_b(i+1)\lambda_c(i+2), \quad a, b, c = x, y, z. \tag{32}$$

(32) contains the $K_{ac}(i, i+2)$ next-to-nearest neighbor two-point correlation that should be eliminated by approximating them with the multiplication of the corresponding two coherence vector elements based on the assumption that the $M_{ac}(i, i+2)$ next-to-nearest neighbor correlation vector elements are zero: $K_{ac}(i, i+2) \approx \lambda_a(i)\lambda_c(i+2)$. Substituting this into (32) leads to the general formula for approximating any nearest neighbor three-point correlation vector element:

$$K_{abc}(i, i+1, i+2) = \langle \hat{\sigma}_a(i)\hat{\sigma}_b(i+1)\hat{\sigma}_c(i+2)\rangle \approx K_{ab}(i, i+1)\lambda_c(i+2) + K_{bc}(i+1, i+2)\lambda_a(i) - \lambda_a(i)\lambda_b(i+1)\lambda_c(i+2), \quad a, b, c = x, y, z. \tag{33}$$

Substituting this into the dynamical equations of nearest neighbor two-point correlations (circled in graph II.a in Table I), the three-point correlations can be eliminated. The method based on this approximation will be called NNPC referring to that besides the coherence vectors it includes only the nearest neighbor pair correlations in the state description of the cell array [17].

The NNPC method is the simplest that is closer to the exact model with the many-body Hamiltonian than the Hartree method. The Hartree method needs $3N$ state variables for state description where $N$ is the number of cells. NNPC requires $3N+9M$ state variables, where $M$ is the number of nearest neighbor pairs among the cells. For a cell line $M=N-1$. Thus the number of state variables scales linearly with the system size for both methods.

The procedure can be generalized. Next-to-nearest neighbor pair correlations and higher than second order correlations can be included and it is also possible to build a model which includes higher order correlations only for those regions where it seems to be necessary.

Since the coherence vector formalism is based on the density matrix description, it is able to model mixed states unlike the state vector description. Dissipation and decoherence can be easily included by adding damping terms to the dynamical equations. This is true for our approximation, as well. Appendix A describes how to add dissipation to the dynamical equations of the coherence vector elements.

## V. Simulation examples

Computer simulations were made to compare NNPC with the Hartree approximation and with the exact model. The comparison was done for the case of quasi-adiabatic switching of a QCA cell line and of a majority gate with unequal input legs. We choose units such that $\hbar$=1 and $E_0$=1. We note that approximating higher and higher order terms puts more and more nonlinear couplings in the differential equations making them numerically more difficult to handle.

### A. Quasi-adiabatic switching of a cell line

The first simulation example is the quasi-adiabatic switching of a line of five cells as shown Fig. 3(a). The first cell is coupled to a driver cell. The tunneling coefficient is gradually[18] lowered (the barriers are raised) as shown in Fig. 3(b). Fig. 3(c) shows the dynamics of the coherence vector coordinates for the five cells coming from NNPC. At the end (when the barriers are high) all the cells align with the driver, that is, at the end $\lambda_z(i) = -P(i) \approx 1$. Fig. 3(d) shows a comparison of the $\lambda_z(2)$ curves corresponding to the Hartree approximation, the NNPC, and the exact model. The inset shows the $\Delta\lambda_z(2)$ deviation from the exact dynamics for the Hartree method (dashed) and NNPC (solid). It is clearly visible that NNPC gives a better match with the exact model than the Hartree approximation does.

Figs. 4(a-b) show the pair correlation vector proper elements for NNPC and the exact model. The $M_{xy}$, $M_{yx}$, $M_{yz}$ and $M_{zy}$ correlation vector proper elements are much smaller than the other five. It can be proved that if the system were exactly in ground state

then they would be zero. NNPC is a qualitative improvement compared to the Hartree approximation since the Hartree approximation does not model intercell correlations at all.

The initial state of the dynamical simulation was the lowest energy stationary state of the NNPC method. The stationary state was found by the multi-dimensional Newton-Raphson method (see Appendix B) using the lowest energy eigenstate of the many-body Hamiltonian of the cell line as starting guess. (The stationary states for the exact model and four our approximation are slightly different. Starting from the initial state of the exact model causes oscillations in the dynamics.) The method works only if (even very small) dissipative terms are added to the dynamical equations. In our simulation the $\tau_{dissip}$ dissipation time constant was $10^7$.

## B. Quasi-adiabatic switching of a majority gate with unequal input legs

For the previous example the Hartree method leads to relatively good results and including correlation terms in the model gives only a quantitative improvement in the dynamics of the polarizations. Next an example, the adiabatic switching of the majority gate with unequal input legs is presented where the results of the Hartree method are even qualitatively wrong.

Before going any further some simplifications must be made in order to reduce the number of state variables and make simulations feasible. The large number of state variables causes a problem even for a majority gate of 9 cells where the coherence vector has $2^{18} - 1 \approx 2.6 \times 10^5$ elements. As it was told in Sec. III, the full set of differential equations for the coherence vector model are equivalent to the many-body Schrödinger equation if the system starts out in pure state and there is no decoherence. Thus for the exact model, the many-body Schrödinger equation will be used, requiring only $2^{10} = 1024$ real state variables. The reduction in the number of state variables is the result of eliminating the degrees of freedom that made it possible to describe mixed states and decoherence. It does not limit our investigation of the role of quantum correlations in the dynamics.

The QCA structure under consideration can be seen in Fig. 5. One of the input legs is only one cell long and coupled to a driver cell with polarization -1. The other two input legs are longer (their length will be denoted by $L$) and they are coupled to drivers with polarization +1. The polarization of the output of the majority gate in ground state (if the barriers are high) is the majority of the polarizations of the input drivers, in this case +1.

The results of the Hartree method are qualitatively wrong for $L$=3 giving an output polarization of -1 as indicated in Fig. 5. The method gives a wrong -1 polarization for cells 4 and 9. (These two cells are circled in the Fig. 5.) The tunneling coefficient is gradually lowered (the barriers are gradually raised) as shown in Fig. 6(a). In Fig. 6(b) the dynamics of the polarizations obtained from the Hartree method can be seen. It is clearly visible that three of the cells settle in the polarization -1 state. Fig. 6(c) shows the exact dynamics. Notice that only one of the cells settles in -1 polarization. The polarization of the gate cell (cell 4) begins to decrease due to the effect of the driver with -1 polarization, however, later it begins to increase and reaches almost +1. (Compare with the dynamics of the gate cell shown in 6(b).)

The phenomenon can be intuitively understood as the results of the competing inputs. Since the leg of the driver with the -1 polarization is shorter, its influence reaches the gate cell first and sets the polarization of the output cell to -1, too. When the other two drivers with the long input legs begin to influence the gate cell, it has already two -1 polarized neighbors thus the drivers are not able to flip the gate cell. The output will be wrong only above a certain difference (L>2) in the length of the input legs. The Hartree method works correctly with a five cell ($L$=1) and a seven cell ($L$=2) gate.

According to our simulations, the NNPC approximation does not model this case correctly, thus further correlation terms must be included in the model. It will be examined which cells of the majority gate can be modeled with the Hartree approach and which must be modeled by a better approximation. The Hartree method assumes that the system is in a product state and the cells are uncorrelated. Thus the parts of the gate where the correlations proper are small can be modeled with the Hartree method, while in the rest of the circuit correlations must be included in the model.

Next it will be checked how large the correlations are in different part of a 9-cell ($L$=3) majority gate. Fig. 7 shows the time dependence of $M_{zz}(1,2)$, $M_{zz}(3,4)$ and $M_{zz}(4,9)$. It can be seen that the latter two (corresponding to correlation in the cross region) are much larger. Thus it seems to be reasonable to include more correlation effects in the five cell cross region. (It must be noted that if all the three inputs are +1 then the correlations in the cross region are much smaller. That is consistent with the fact that the Hartree method, that does not include correlations at all, works well in this case.)

One possibility is to include all the two-point, three-point, four-point, and five-point correlations of the cross region in the model while handling all the other cells with the Hartree approximation. To do that certain simplifications are needed, since the number of state variables for the set of differential equation for the coherence vector elements are very large (1035) even for the 9-cell gate. (Notice that this number is still much smaller than the one obtained for the full set of differential equations having all the correlations.).

The simplification can be based on recognizing that in the coherent case the previous model is equivalent to a set of coupled Schrödinger equations:

$$i\hbar\frac{d\Psi_{cross}}{dt} = \hat{H}_{cross}\Psi_{cross}\text{, and} \tag{34}$$

$$i\hbar\frac{d\Psi_k}{dt} = \hat{H}_k\Psi_k \text{ for } k\text{=1,2,6.7.} \tag{35}$$

(34) describes the time evolution of the state of the cross region (cells 3,4,5,8, and 9), while (35) single-cell Schrödinger equations describe the time evolution of the state of the remaining cells.

The five-cell many-body Hamiltonian for the cross is

$$\begin{aligned}\hat{H}_{cross} = -\gamma \sum_{i\,=\,3,\,4,\,5,\,8,\,9} \hat{\sigma}_x(i) - \\ \frac{E_0}{2}(\hat{\sigma}_z(3)\hat{\sigma}_z(4) + \hat{\sigma}_z(4)\hat{\sigma}_z(5) + \hat{\sigma}_z(9)\hat{\sigma}_z(4) + \hat{\sigma}_z(8)\hat{\sigma}_z(4)) - \\ \frac{E_1}{2}(\hat{\sigma}_z(3)\hat{\sigma}_z(8) + \hat{\sigma}_z(3)\hat{\sigma}_z(9) + \hat{\sigma}_z(5)\hat{\sigma}_z(8) + \hat{\sigma}_z(5)\hat{\sigma}_z(9)) - \\ \frac{E_0}{2}(\hat{\sigma}_z(3)P(2) + \hat{\sigma}_z(8)P(7) + \hat{\sigma}_z(5)P_{driver}(5)),\end{aligned} \tag{36}$$

while the single-cell Hamiltonians are

$$\hat{H}_k = -\gamma\hat{\sigma}_x(k) + \frac{\Sigma_k}{2}\hat{\sigma}_z(k)\,. \tag{37}$$

(36) is coupled to the neighboring cell through the $P(2)$, $P(7)$ and $P_{driver}(5)$ polarizations while (36) is coupled to the environment through $\Sigma_i$. $E_1$=-0.18$E_0$ describes the interaction between diagonal neighbors. The negative sign indicates that they tend to anti-align. The relative strength of the diagonal interaction is computed from geometrical considerations.

If there is no dissipation, the coherence vector description with the approximation that includes correlations only in the cross region would give the same dynamics for the system as the (34-35) system of coupled Schrödinger equations do. By neglecting the degrees of freedom coming from the ability of the coherence vector description to model mixed states, the number of real state variables are reduced to 80.

Fig. 8(a) shows the dynamics of the polarizations for the 9-cell gate ($L=3$). It now gives the correct polarization for the output cell. (Compare with Fig. 6.) Fig. 8(b) shows the dynamics of $M_{zz}(3,4)$ and $M_{zz}(4,9)$ in the cross region. Comparison with Fig. 7(b) indicates that large part of the correlations proper are restored. It is worthwhile to try how long the difference between the input legs can be before the method breaks down and gives the wrong answer. Simulations show that this approximation gives the correct output for $L<40$. (Notice the large improvement compared to the Hartree method that worked correctly for $L<3$.)

Originally it was thought that the Hartree approach fails for the majority gate since, because of the inequality of the input legs, the influence of one of the drivers reaches the gate cell before the other two. Our findings support the idea that what caused the Hartree approach to fail was its inability of modeling the correlations in the cross region. (Notice that in Fig. 8(a) the polarization of the gate cell (cell 4) decreases in the beginning, similarly to the results of the exact method shown in Fig. 6(c). Thus with both models the influence of the driver with -1 polarization reaches the gate cell before the influence of the other two inputs does.) If all the three inputs are +1 then even the Hartree method gives the correct dynamics consistently with the previous remark about the role of correlations since in this case the correlations in the cross region are much smaller.

# VI. Conclusions

An intermediate model between the Hartree approximation and the exact method was constructed to describe the dynamics of QCA cell arrays. It is based on the truncation of the system of dynamical equations obtained from the coherence vector formalism. By choosing the point of truncation it is possible to include correlation effects to the desired order in the dynamics. The Nearest Neighbor Pair Correlation (NNPC) model kept all the nearest neighbor two-point correlations while approximating the three-point correlations and the further than nearest neighbor two-point correlations. Through the example of the majority gate with unequal input legs it was also shown how to construct an approximation where the correlations are included fully only in a certain part of the circuit while other parts are modeled by dynamical equations using the intercellular Hartree approximation. The method corrects the qualitatively wrong results of the Hartree method in determining the output for the gate. The usefulness of these models can be summarized as follows. (1) They *quantitatively* improve the dynamics of the single-cell coherence vectors compared to the Hartree model. (2) They represent a *qualitative* improvement since they give the (approximate) dynamics of the correlation while the Hartree model does not give information on correlation. (3) These approximate models help understating which quantum degrees of freedom are important from the point of view of the dynamics.

## Acknowledgement

The authors would like to thank John Timler and György Csaba for stimulating discussions. This work was supported by the Office of Naval Research MURI program.

## Appendix A: Including dissipation in the dynamical equations

The model presented in the previous subsections describes the unitary time evolution of the cell line based on the (13) dynamical equation of the coherence vector. Inserting damping terms in the differential equations [15] for the coherence vector and correlation vector elements, dissipation can also be included in the dynamics. The (14) differential equation for the single-cell coherence vector changes in the following way:

$$\frac{d}{dt}\vec{\lambda}(i)\Big|_{diss} - \frac{d}{dt}\vec{\lambda}(i)\Big|_{ndiss} = -\frac{1}{\tau_{dissip}}(\vec{\lambda}(i) - \vec{\eta}(i)) \ . \tag{38}$$

where *diss* and *ndiss* stand for dissipative and non-dissipative. $\frac{1}{\tau_{dissip}}$ describes the dissipation rate. Vector $\vec{\eta}_i$ accounts for the fact that the dissipation drives the coherence vector elements to non-zero values.

The (16) differential equation for a correlation vector changes in the following way:

$$\frac{d}{dt}\vec{K}(i,j)\Big|_{diss} - \frac{d}{dt}\vec{K}(i,j)\Big|_{ndiss} = -\frac{2}{\tau_{dissip}}\left(\vec{K}(i,j) - \frac{\vec{\eta}(i) \otimes \vec{\lambda}(j) + \vec{\lambda}(i) \otimes \vec{\eta}(j)}{2}\right). \tag{39}$$

The changes for equations of higher order correlations can be found in [15].

There are several possibilities to choose the $\vec{\eta}(i)$ vector depending on what kind of model of dissipation is used. One possibility is the following. The instantaneous ground state with no dissipation according to the Hartree model is given by (27-28). $\vec{\eta}(i)$ can be chosen as

$$\vec{\eta}_i(i) = \frac{1}{\tau_{dissip}}\vec{\lambda}_{ss}(i) = -\frac{1}{\tau_{dissip}}\frac{\vec{\Gamma}_i}{|\vec{\Gamma}_i|} \ . \tag{40}$$

Modeling the dissipation this way, describes the relaxation of the coherence vector towards $\vec{\lambda}_{ss}$. If $\frac{1}{\tau_{dissip}}$ is large then the system follows closely the instantaneous ground state of the Hartree model.

## Appendix B: Finding the stationary state of the dynamical equations

The stationary states of NNPC can be obtained taking all the time derivatives zero in the dynamical equations and solving for the coherence vector and correlation vector elements. The dynamics of the system can be written in the general form:

$$\frac{d}{dt}\vec{\Lambda} = F(\vec{\Lambda}), \tag{41}$$

where $F(\vec{\Lambda})$ is a vector-valued function of the vector variable $\vec{\Lambda}$. (The (13) differential equations giving the exact dynamics for the coherence vector are linear, however, the NNPC method uses nonlinear terms to approximate higher order correlations.) The stationary solution of (41) can be obtained from

$$0 = F(\vec{\Lambda}_{stat}). \tag{42}$$

We used the multi-dimensional Newton-Raphson method to find $\vec{\Lambda}_{stat}$. It converges very fast since $F(\vec{\Lambda})$ contains mostly linear terms, except for the terms approximating the higher order correlations.

The multi-dimensional Newton-Raphson method is based on the linearization of $F(\vec{\Lambda})$ around an initial guess, $\vec{\Lambda}_{ini}$. The next guess, $\vec{\Lambda}_{next}$, will be the vector that makes the linearized function zero. The linearization of $F(\vec{\Lambda})$ is

$$F(\vec{\Lambda}) - F(\vec{\Lambda}_{ini}) \approx J(\vec{\Lambda}_{ini})(\vec{\Lambda} - \vec{\Lambda}_{ini}). \tag{43}$$

Here $J(\vec{\Lambda}_{ini})$ is the Jacobian of $F(\vec{\Lambda})$ at $\vec{\Lambda}_{ini}$. Since we are looking for the zero of $F(\vec{\Lambda})$, the following equation must be solved for $\vec{\Lambda}_{next}$:

$$-F(\vec{\Lambda}_{ini}) = J(\vec{\Lambda}_{ini})(\vec{\Lambda}_{next} - \vec{\Lambda}_{ini}). \tag{44}$$

The solution is

$$\vec{\Lambda}_{next} = \vec{\Lambda}_{ini} - J^{-1}(\vec{\Lambda}_{ini})F(\vec{\Lambda}_{ini}). \tag{45}$$

This gives the next guess from the previous guess. Notice that the Jacobian must be invertible since (45) explicitly contains its inverse. The Jacobian is singular if there is no dissipation, thus adding (even very small) damping terms to the equations is necessary to find the stationary state. It is reasonable to determine the Jacobian analytically instead using numerical differentiation in order to increase the computational speed and the accuracy as well.

## Tables

TABLE I. The hierarchy of the dynamical equations for the coherence vector elements for a cell line. The first three levels are shown: dynamics of the single cell coherence vectors (graph I), nearest neighbor two-point correlation vectors (graph II.a), further-than-nearest neighbor two-point correlation vectors (graph II.b), nearest neighbor three-point correlation vectors (graph III). The graphs are indicating which variables are in the dynamical equations of a particular coherence vector element. The dashed and dashed-dotted lines show where the Hartree method and the NNPC approximation (see text) truncate the hierarchy.

## Figure Captions

FIGURE 1. Schematic of the basic four-site semiconductor QCA cell. (a) The geometry of the cell. The tunneling energy between two sites (quantum dots) is determined by the heights of the potential barrier between them. (b) Coulombic repulsion causes the two electrons to occupy antipodal sites within the cell. These two bistable states result in cell polarization of $P$=+1 and $P$=-1. (c) Nonlinear cell-to-cell response function of the basic four-site cells. Cell 1 is a driver cell with fixed charge density. In equilibrium the polarization of cell 2 is determined by the polarization of cell 1. The plot shows the polarization $P_2$ induced in cell 2 by the polarization of its neighbor, $P_1$. The solid line corresponds to antiparallel spins, and the dotted line to parallel spins. The two are nearly degenerate especially for significantly large values of $P_1$.

FIGURE 2. The steps of the quasi-adiabatic switching are the following: (1) before applying the new input, the height of the interdot barriers are lowered thus the cell have no more two distinct polarization states, $P$=+1 and $P$=-1. (2) Then the new input can be given to the array. (3) While raising the barrier height, the QCA array will settle in its new ground state. The quasi-adiabaticity of the switching means that the system is very close to its ground state during the whole process. It does not get to excited state after setting the new input, as it happened in the case of non-adiabatic switching. Since the system does not get to an excited state from the ground state the dissipation decreased a lot.

FIGURE 3. Quasi-adiabatic switching of a line of five cells. The barriers are gradually lowered while the driver has constant -1 polarization. The five cells follow the polarization of the driver. (a) The arrangement of the five cells and a driver, (b) the dynamics of the interdot tunneling energy, (c) the elements of the three coherence vectors as the function of time for the NNPC approximation, (d) $\lambda_z(2)$ as the function of time for the Hartree approximation (dashed), NNPC (solid), and the exact model (solid). The inset shows the $\Delta\lambda_z(2)=\lambda_z(2)-\lambda_{z,exact}(2)$ deviation from the exact dynamics for the Hartree method (dashed) and NNPC (solid). NNPC gives a result closer to the exact one than the Hartree approximation does.

FIGURE 4. Quasi-adiabatic switching of five cells. The barriers are gradually raised while the driver has constant -1 polarization. The nearest neighbor correlation vector proper elements for the (a) the NNPC approximation and (b) the exact method. The $M_{xy}$, $M_{yx}$, $M_{yz}$ and $M_{zy}$ correlation vector proper elements are much smaller than the other five, thus they are multiplied by 100.

FIGURE 5. 9-cell (*L*=3) majority gate with unequal input legs. At the end of the quasi-adiabatic switching process, when the barriers are high, the output polarization of the majority gate should be the majority polarization of the inputs. When modeled by the Hartree method, the polarizations of the cells #4 and #9 (circled) are determined incorrectly. In the correct ground state all the cells have +1 polarization except for cell #5 which has -1.

FIGURE 6. Quasi-adiabatic switching of a 9-cell majority gate (*L*=3). (a) The time dependence of the tunneling energy. The barriers are gradually raised. (b) The cell polarizations as the function of time for the Hartree method and (c) for the exact model. In both (b) and (c) the curves corresponding to cell 4 (gate cell), cell 9 (output cell) and cell 5 are labeled.

FIGURE 7. Dynamics of the two-point correlations proper during the quasi-adiabatic switching of a 9-cell majority gate (*L*=3). $M_{zz}(1,2)$ (dashed-dotted), $M_{zz}(3,4)$ (solid) and $M_{zz}(4,9)$ (dashed) are shown. The correlations are much larger in the cross region than away from it.

FIGURE 8. Quasi-adiabatic switching of a 9-cell majority gate (*L*=3). All the correlations are included in the five-cell cross region while outside this region a Hartree description is used. (a) Dynamics of the polarizations. The curves corresponding to cell 4 (gate cell), cell 9 (output cell) and cell 5 are labeled.

FIGURE 9. (b) Dynamics of the two-point correlations proper $M_{zz}(3,4)$ (solid) and $M_{zz}(4,9)$ (dashed). Part of the correlations is restored in the cross region. Compare with Fig. 7.

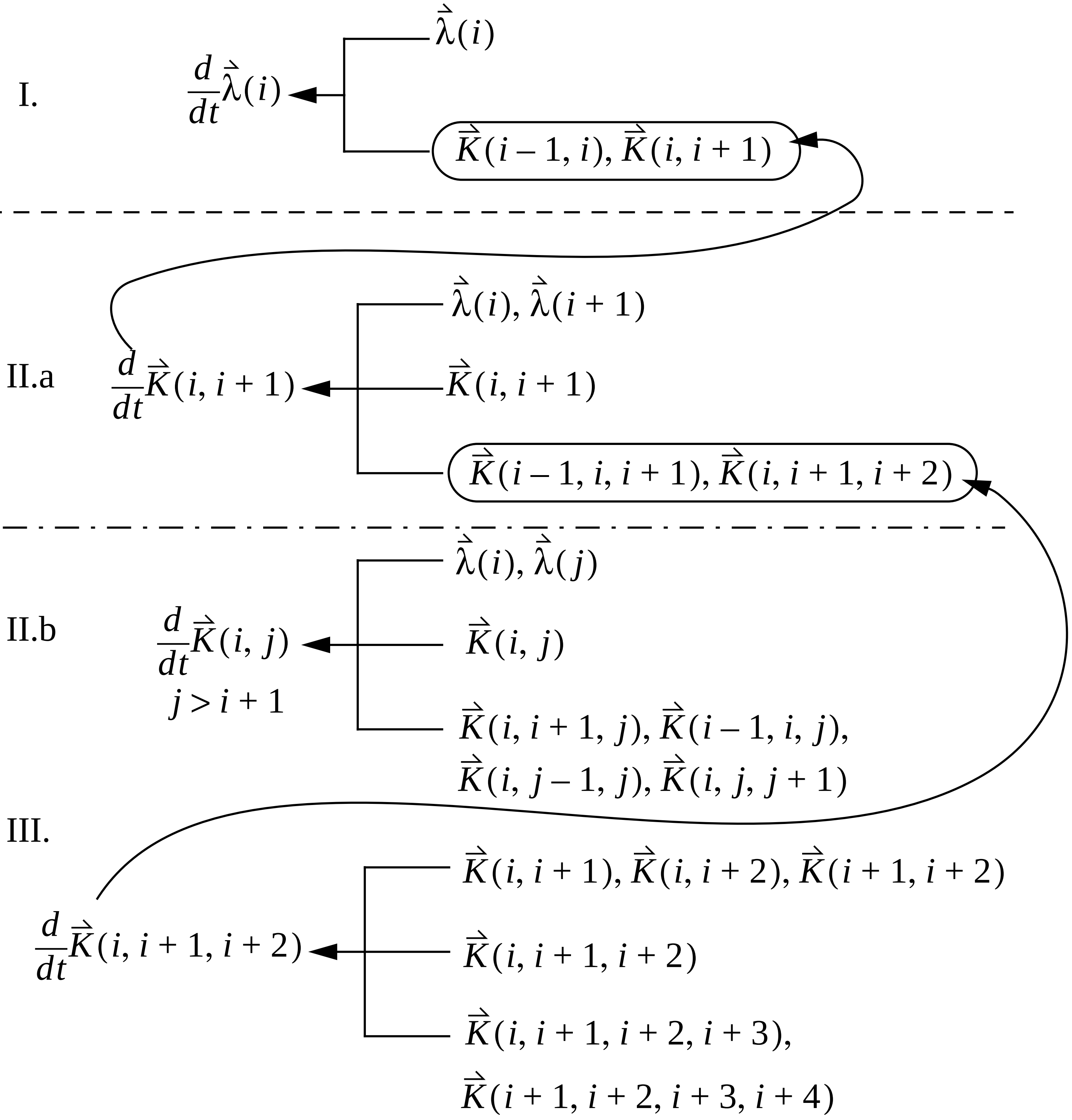
I.
$\frac{d}{dt}\vec{\lambda}(i)$
$\vec{\lambda}(i)$
$\vec{K}(i-1, i), \vec{K}(i, i+1)$
II.a
$\frac{d}{dt}\vec{K}(i, i+1)$
$\vec{\lambda}(i), \vec{\lambda}(i+1)$
$\vec{K}(i, i+1)$
$\vec{K}(i-1, i, i+1), \vec{K}(i, i+1, i+2)$
II.b
$\frac{d}{dt}\vec{K}(i, j)$
$j > i+1$
$\vec{\lambda}(i), \vec{\lambda}(j)$
$\vec{K}(i, j)$
$\vec{K}(i, i+1, j), \vec{K}(i-1, i, j),$
$\vec{K}(i, j-1, j), \vec{K}(i, j, j+1)$
III.
$\frac{d}{dt}\vec{K}(i, i+1, i+2)$
$\vec{K}(i, i+1), \vec{K}(i, i+2), \vec{K}(i+1, i+2)$
$\vec{K}(i, i+1, i+2)$
$\vec{K}(i, i+1, i+2, i+3),$
$\vec{K}(i+1, i+2, i+3, i+4)$

Table I.

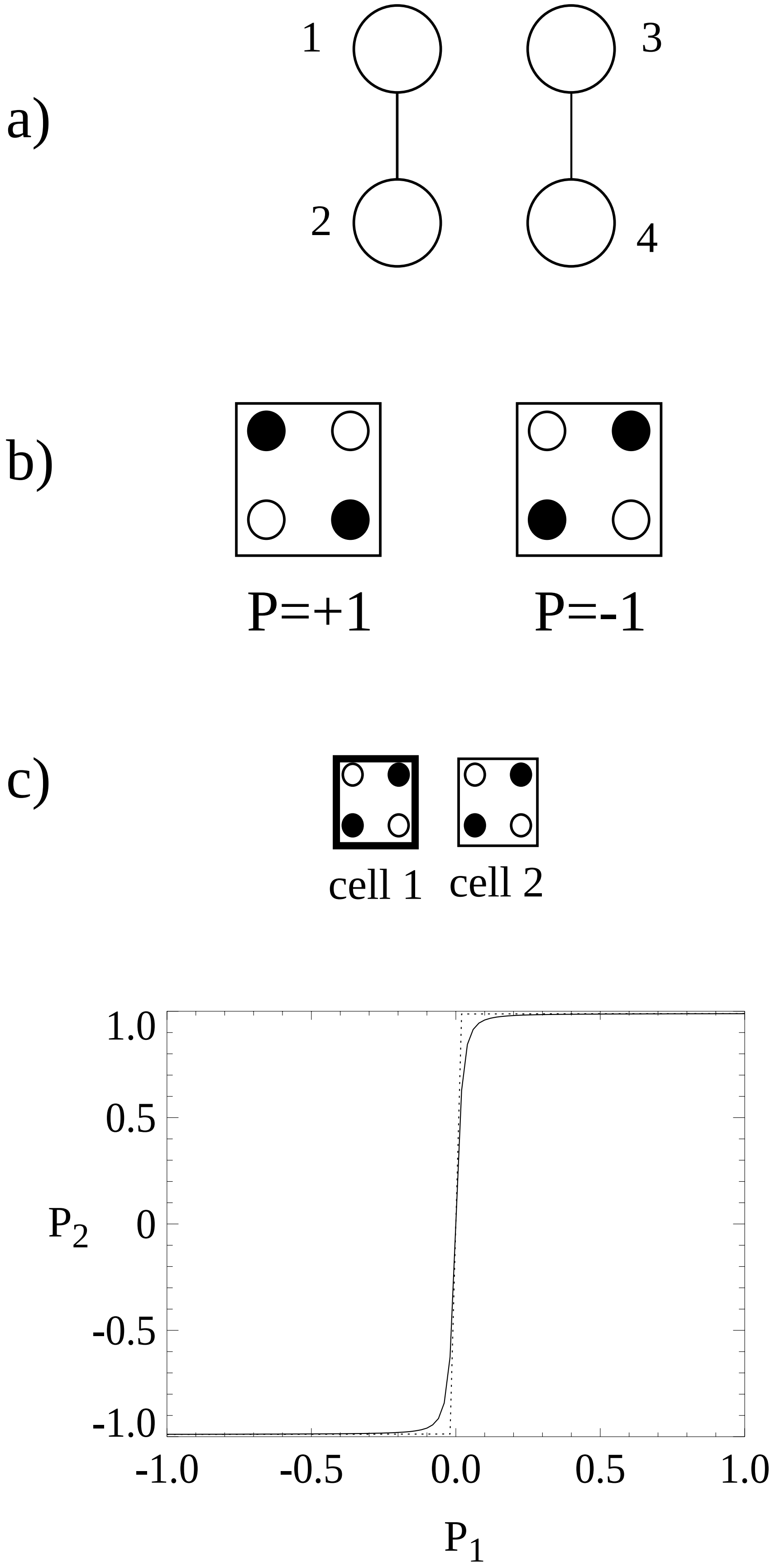

Figure 1

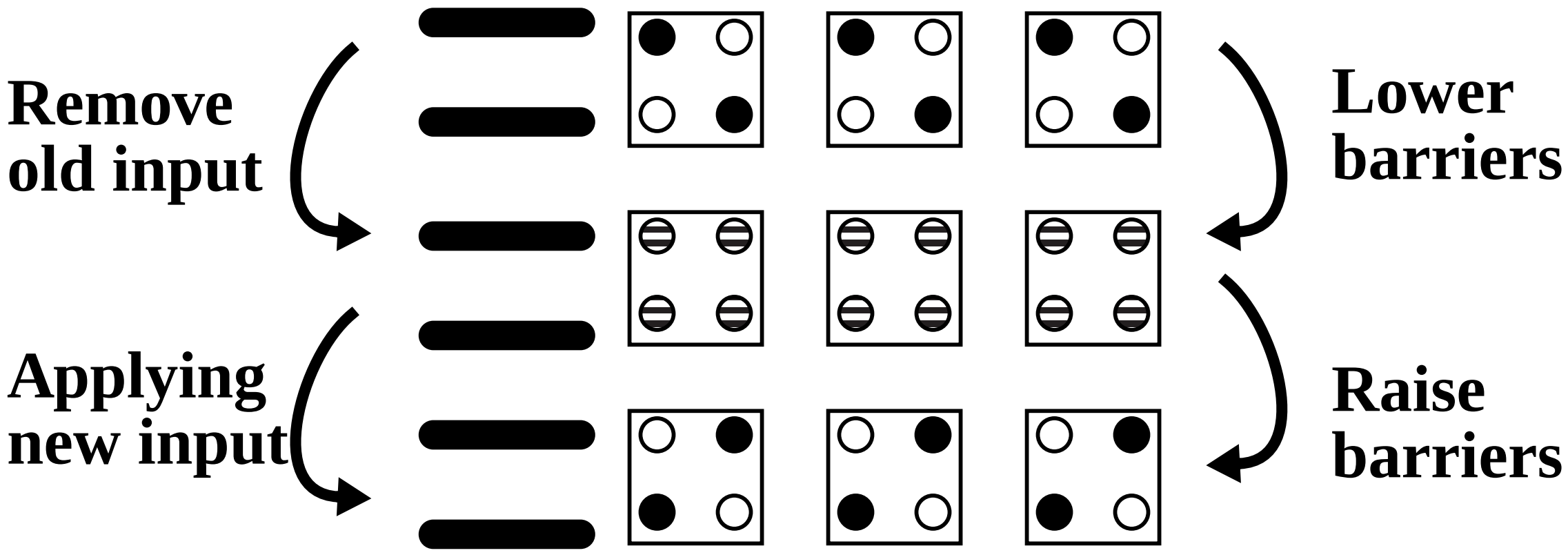

Figure 2

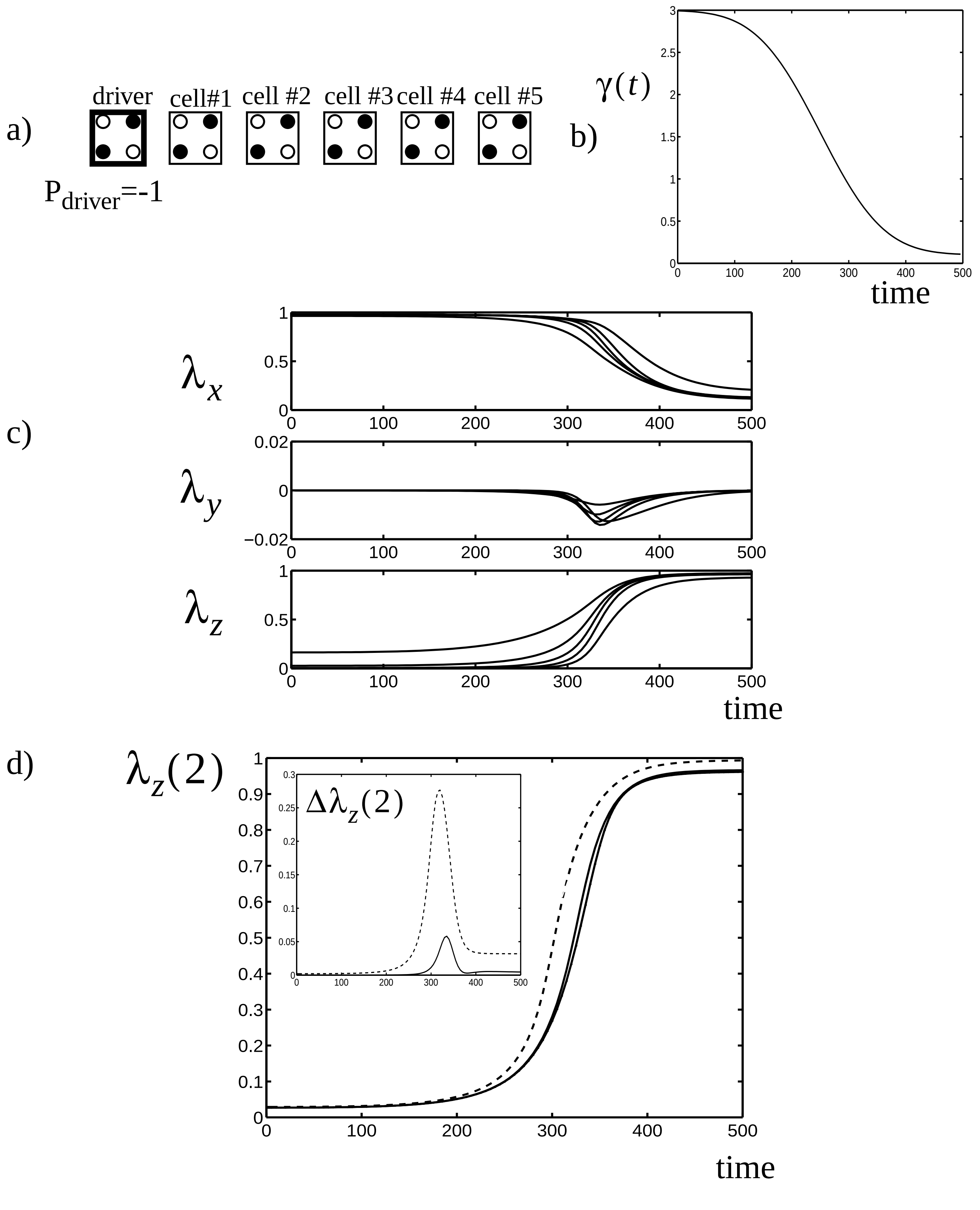

Figure 3

a)

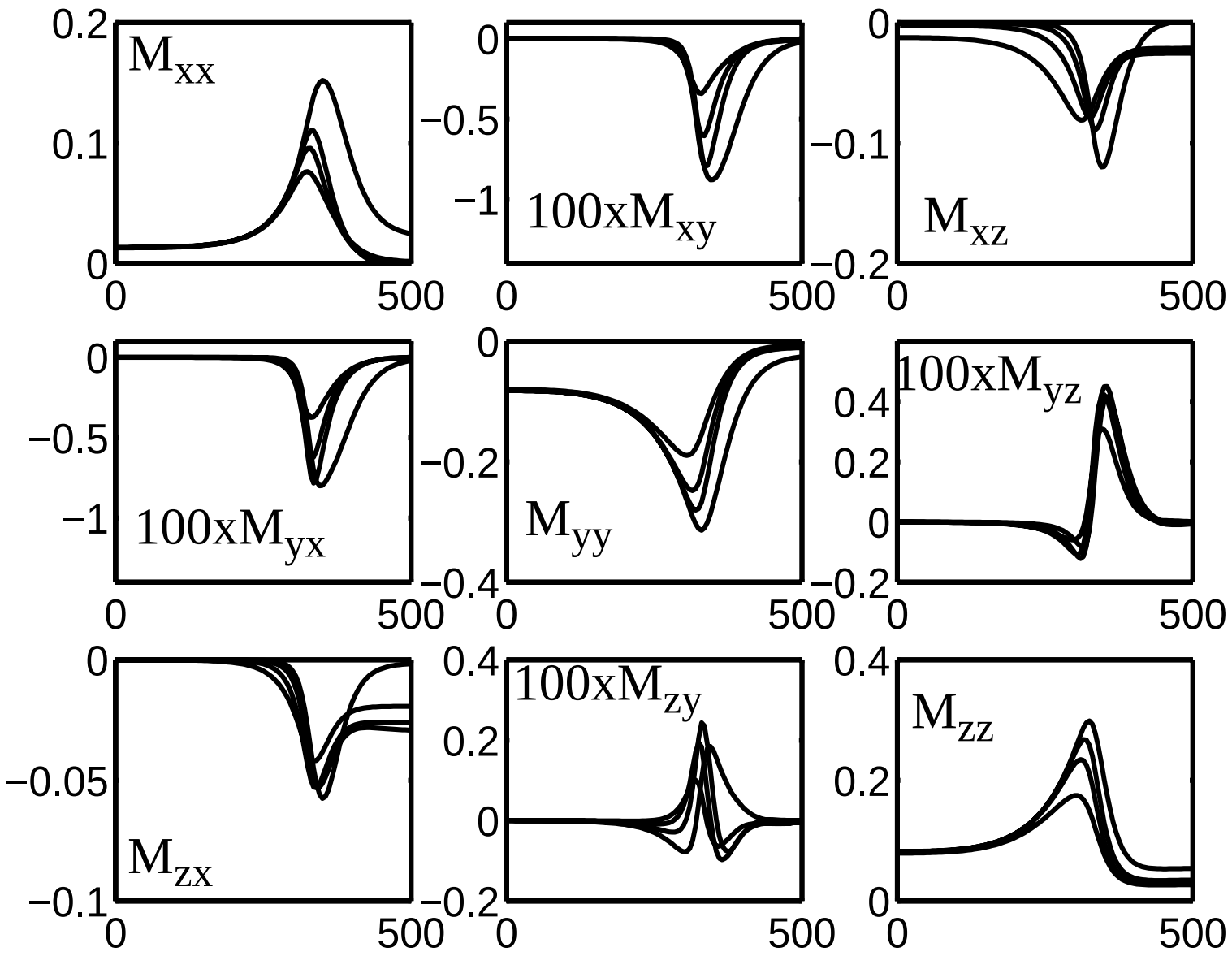

b)

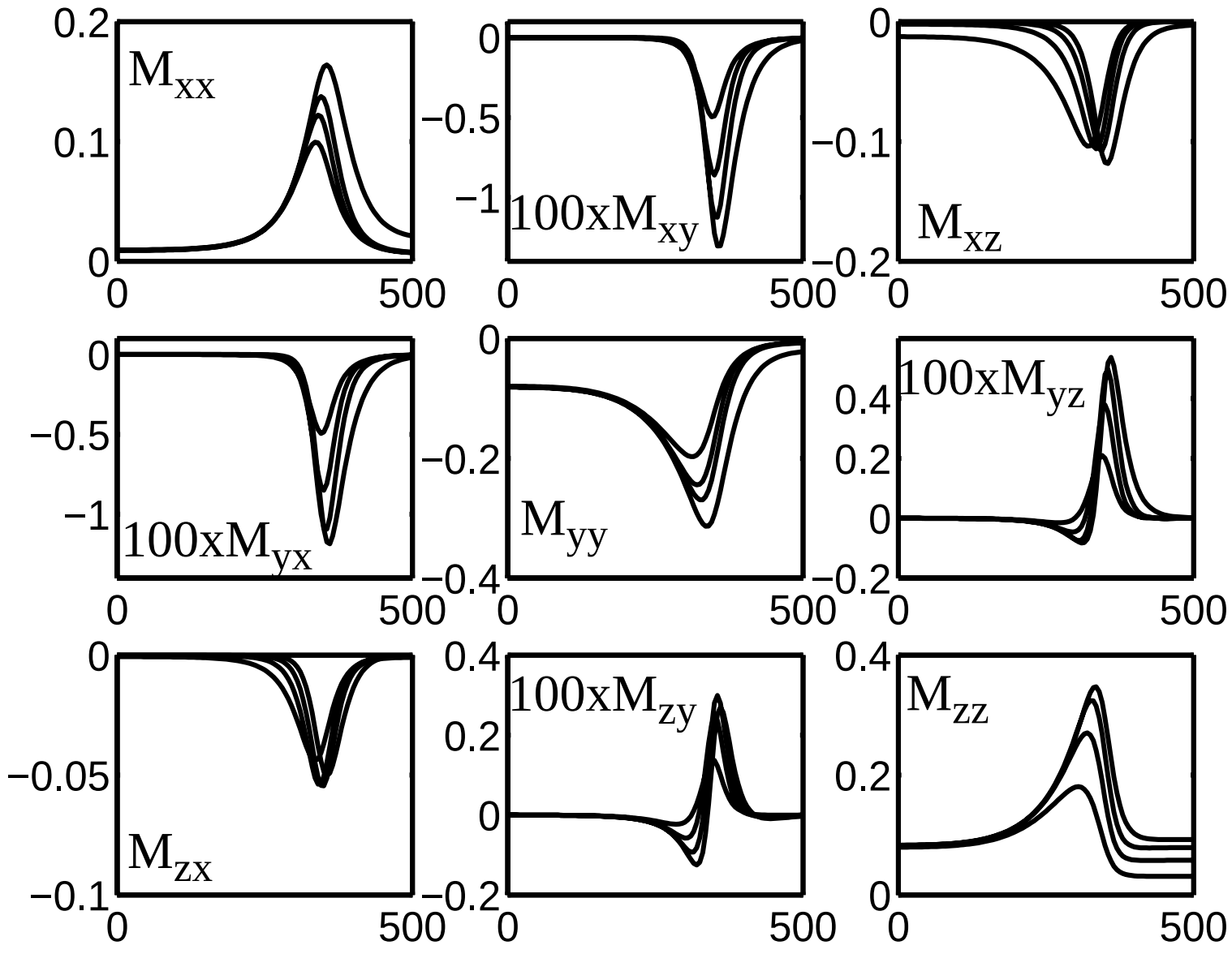

Figure 4

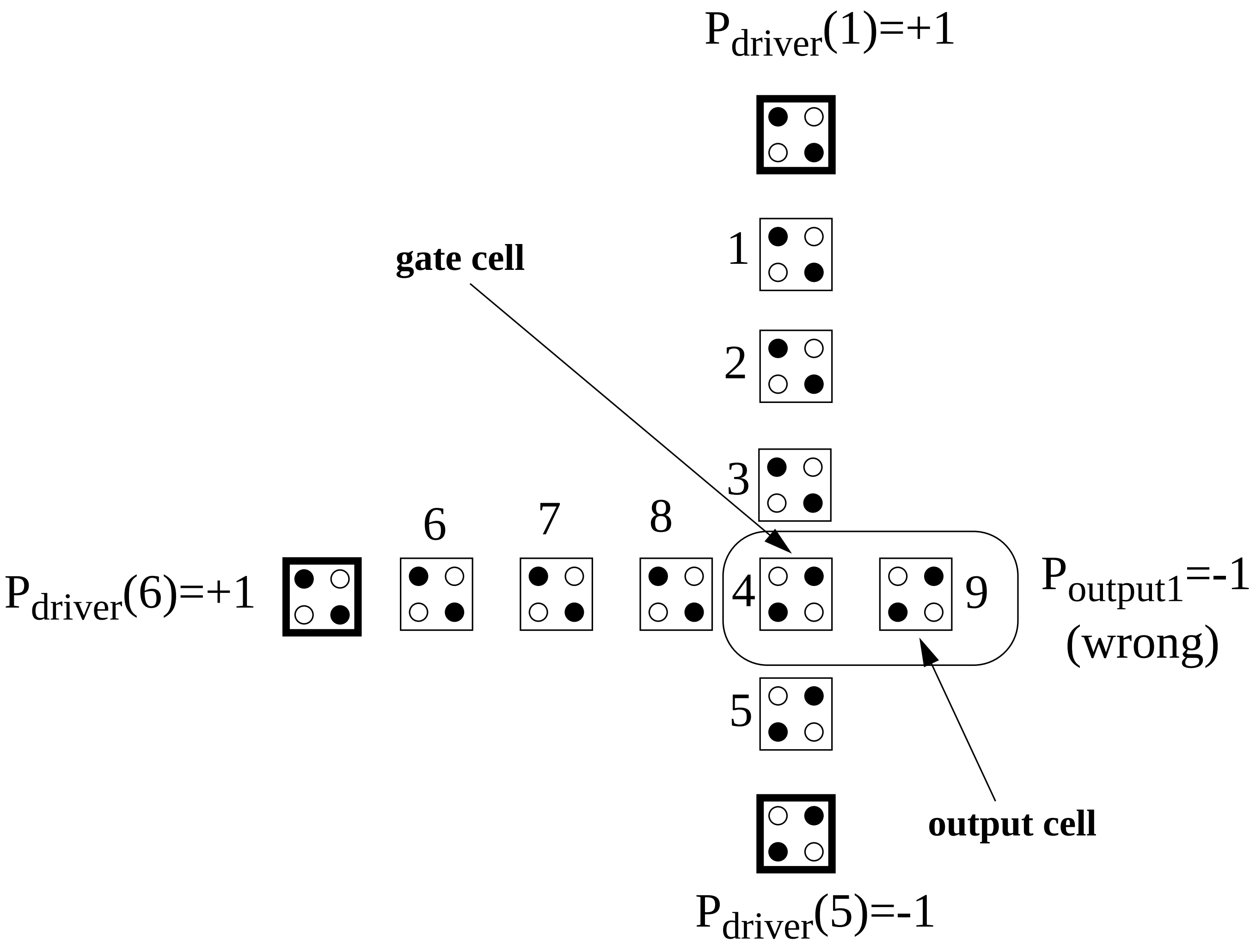

Figure 5

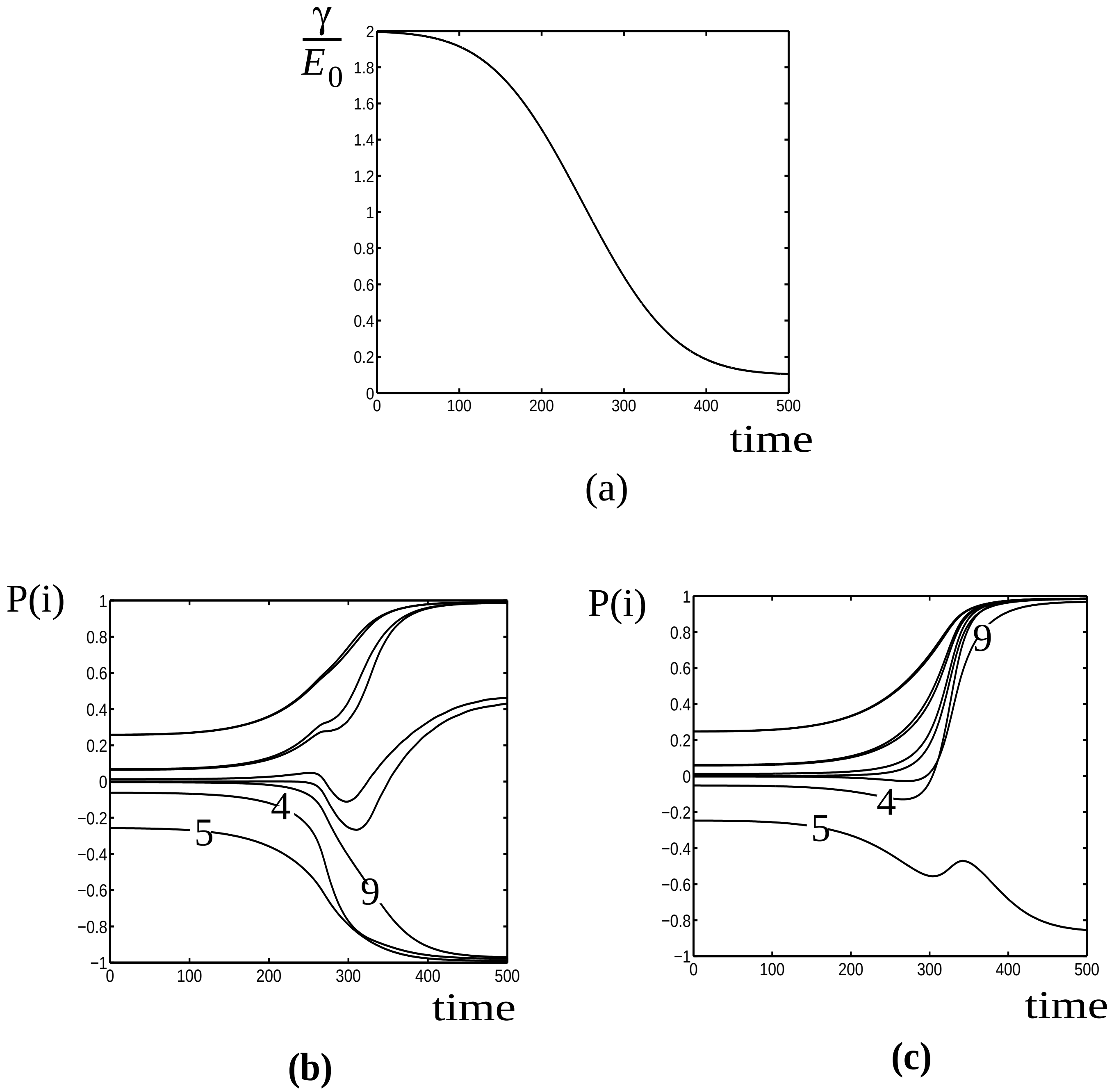

Figure 6

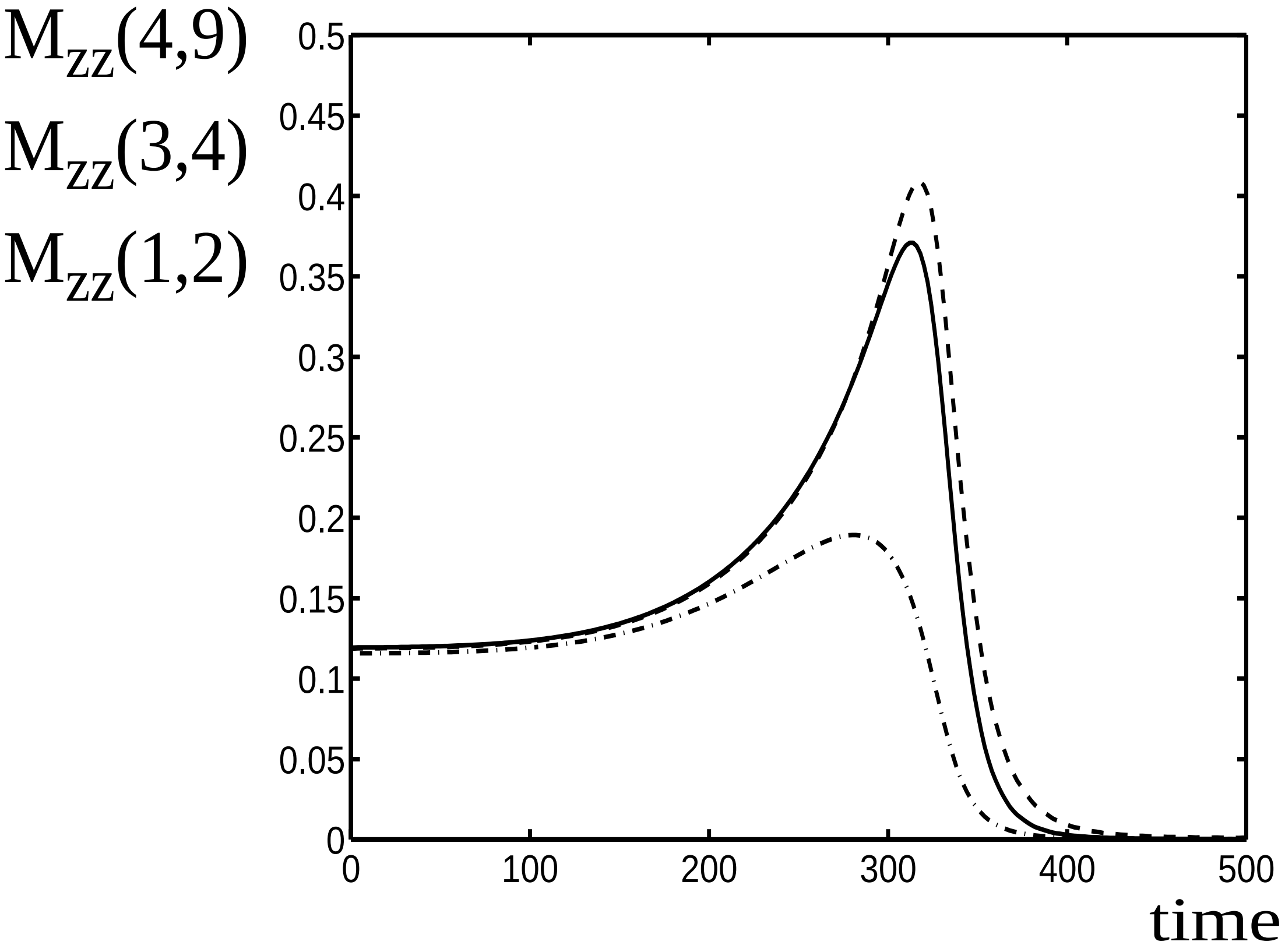

Figure 7

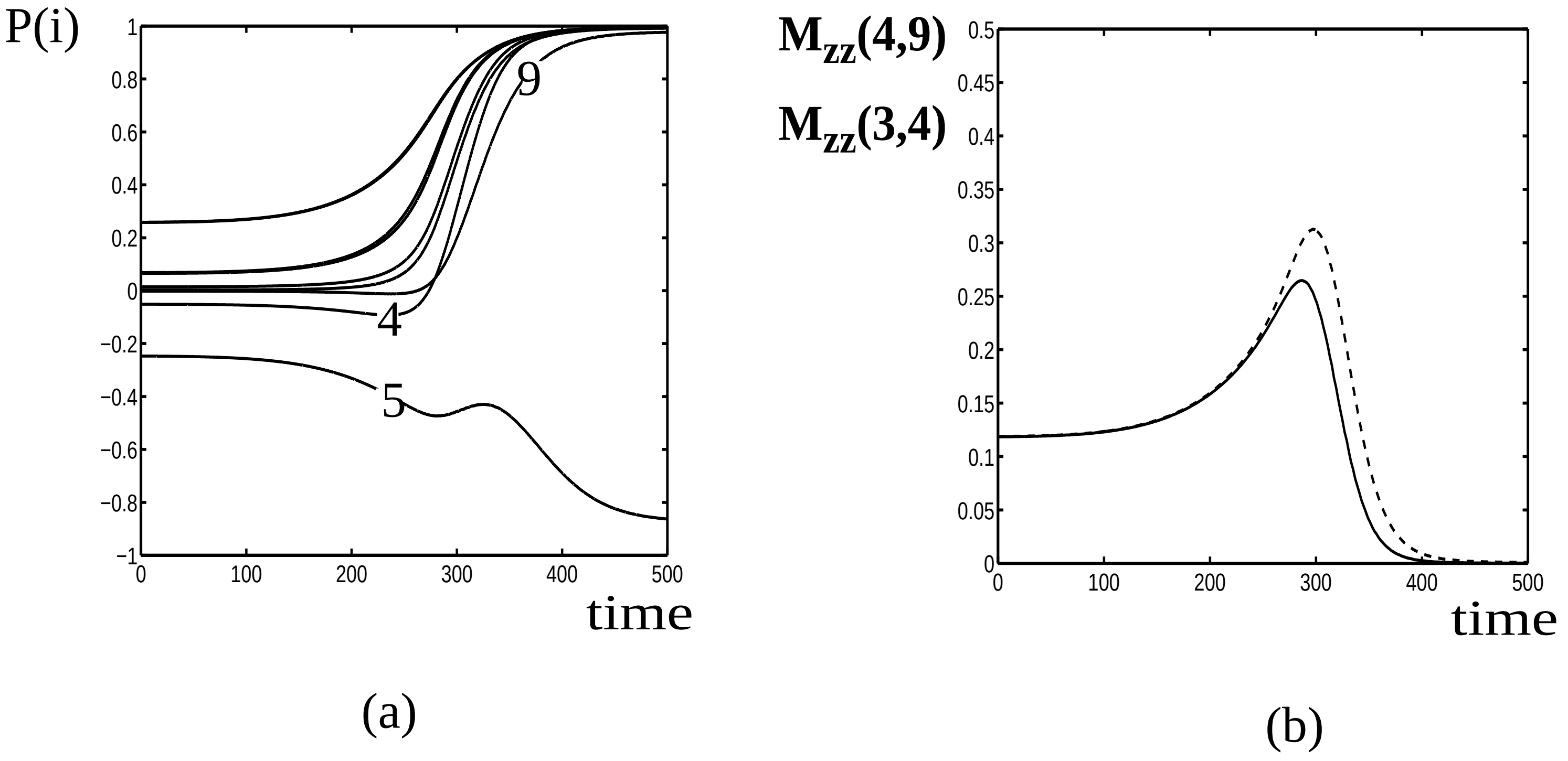

Figure 8